\documentclass[12pt,preprint]{aastex}

\shorttitle{Solar source of the impulsive electron events}
\shortauthors{Li et al.}

\begin{document}

%% LaTeX will automatically break titles if they run longer than
%% one line. However, you may use \\ to force a line break if
%% you desire.

\title{Coronal jets, magnetic topologies, and the production of
interplanetary electron streams}

%% Use \author, \affil, and the \and command to format
%% author and affiliation information.
%% Note that \email has replaced the old \authoremail command
%% from AASTeX v4.0. You can use \email to mark an email address
%% anywhere in the paper, not just in the front matter.
%% As in the title, use \\ to force line breaks.

\author{C. Li\altaffilmark{1}, S. A. Matthews\altaffilmark{1},
L. van Driel-Gesztelyi\altaffilmark{1,2,3}, J. Sun\altaffilmark{1},
and C. J. Owen\altaffilmark{1}}

\altaffiltext{1}{Mullard Space Science Laboratory, University
College London, Dorking, Surrey RH5 6NT, UK}
\email{cl2@mssl.ucl.ac.uk}
\altaffiltext{2}{Observatoire de Paris, LESIA, UMR 8109 (CNRS),
92195 Meudon Cedex, France}
\altaffiltext{3}{Konkoly Observatory of Hungarian Academy of
Sciences, Budapest, Hungary}

%% Mark off your abstract in the ``abstract'' environment. In the
%% manuscript style, abstract will output a Received/Accepted line
%% after the title and affiliation information. No date will appear
%% since the author does not have this information. The dates will
%% be filled in by the editorial office after submission.

\begin{abstract}

We investigate the acceleration source of the impulsive solar
energetic particle (SEP) events on 2007 January 24. Combining the
in-situ electron measurements and remote-sensing solar observations,
as well as the calculated magnetic fields obtained from a
potential-field source-surface (PFSS) model, we demonstrate that the
jets associated with the hard X-ray (HXR) flares and type III radio
bursts, rather than the slow partial coronal mass ejections (CMEs),
are closely related to the production of interpanetary electron
streams. The jets, originated from the well-connected active region
(AR 10939) whose magnetic polarity structure favors the eruption,
are observed to be forming in a coronal site, extending to a few
solar radius, and having a good temporal correlation with the
electron solar release. The open-field lines nearby to the jet site
are rooted in negative polarities, along which energetic particles
escape from the flaring AR to the near-Earth space, consistent with
the in-situ electron pitch angle distribution (PAD). The analysis
enable us to propose a coronal magnetic topology relating the
impulsive SEP events to their solar source.

\end{abstract}


\section{INTRODUCTION}

A class of solar energetic particle (SEP) events, which are
associated with ``impulsive" X-ray flares and type-III radio bursts,
show distinct enhancement of $e/p$ ratio and in some cases high-Z
elemental abundance (e.g., $^{3}$He and Fe). Such energetic
electrons and heavy ions are believed to be originated from flaring
active regions (ARs), separating from large coronal mass ejection
(CME) and shock-related ``gradual" events \citep{rea99,kal03,mas07}.
Apart from the irreducible requirement of a western hemisphere AR
where magnetic and plasma processes preferentially energize and
release particles \citep{mas09}, we still do not clearly identify
the solar source of the impulsive SEP events.

Recently, \citet{wan06} and \citet{nit08} proposed that jets seen in
extreme-ultraviolet (EUV) and sometimes in white-light (WL) images
are closely related to the $^{3}$He-rich SEP events. The jets may be
explained as expanding loops reconnecting with large-scale unipolar
magnetic fields \citep{shi94}, which are open to interplanetary
space for energetic particles to be observed at 1 AU. However,
\citet{kah01} and \citet{pic06} found some unusual cases of
$^{3}$He-rich SEP events are associated with narrow and fast CMEs,
making the source identification more complex. A possibility is
suggested that particles are accelerated close to the jets or
plasmoids, which move upward from magnetic reconnection sites and
might appear as faint and narrow CMEs in coronagraphs \citep{pic06}.
These recent progresses in identifying the solar source still leave
unsolved questions: Do the jets or CMEs in association with the
impulsive SEP events play a critical role in particle acceleration
\citep{mas09}? How does the magnetic topology in the source region
relate solar activities to the in-situ SEP dynamics \citep{li10}? To
answer the questions, careful comparison of in-situ particle
measurements and remote-sensing solar observations, as well as the
modeling of coronal magnetic fields should be carried out.

In this study we present a cross-disciplinary investigation of the
impulsive SEP events observed on 2007 January 24, by using energetic
electron data to constrain the in-situ SEP dynamics (for instance,
timing and distribution), which are much more accurate than with
low-energy ion data, and the multi-wavelength imaging data to detect
the details of the solar activities. We also apply the velocity map,
magnetogram and the potential-field extrapolation to interpret the
magnetic morphology in the source region. Our purpose is to clarify
the links between jet eruption, coronal magnetic topology, and the
production of interplanetary electron streams.

\section{IN-SITU MEASUREMENTS}

Two successive electron events were observed on 2007 January 24 by
the WIND three-dimensional Plasma and Energetic Particles instrument
\citep{lin95} and the Electron, Proton, and Alpha Monitor
\citep{gol98} on board Advanced Composition Explorer (ACE), both are
orbiting the Sun-Earth L1 liberation point. The WIND/3DP provides
electron measurements with the electrostatic analyzers (EESAs) from
$\sim$ 0.5 to 28 keV, and with the solid-state telescopes (SSTs) from
27 to $\sim$ 300 keV which suffered a data gap during the events.
Compensating for the data gap, the ACE/EPAM measures electrons with
the Composition Aperture (CA), whose look direction is oriented
60$^{\circ}$ from the spacecraft spin axis and so is known as the
CA60 telescope, in the energy range of 38 $-$ 315 keV.

Figure 1 shows the electron intensity profiles observed from
$\sim$ 1 to $\sim$ 300 keV by the WIND/3DP and ACE/EPAM instruments,
respectively shown in Panel (a) and (b), together with the GOES soft
X-ray (SXR) $1 - 8 \ \rm {\AA}$ light curve shown in Panel (c). Two
successive flux enhancements, marked as event 1 and event 2, were
recorded. Both events show clear velocity dispersion (injection and
peak times are later for lower energies), which are the typical
identifications of their solar origin. They correspond to the
impulsive B5.5 and B7.3 flares, respectively, which were erupted
from AR 10939 in the west hemisphere. When the flares occurred, the
position of this AR was roughly at S06W60 nearby to the footpoints
of interplanetary magnetic field (IMF) lines connecting the Sun to
the near-Earth space, facilitating energetic particles escape to the
in-situ spacecraft. The larger B9.1 flare, even though accompanied
by a relatively speaking fast (projection speed of 785 km/s) and
wide (angular width of $147^{\circ}$) CME, was produced by a eastern
AR located at S04E110, too far from the well-connected region,
therefore was not related to a SEP event.

Since there were severe data gaps of solar imaging observations in
the time window of event 1. On the other hand, the two events show
many similarities between them (see discussion in Section 5).
Therefore, we mainly focus on the event 2 in the following study.
To determine the solar release time, assuming electrons travel along
the IMF lines at a speed of $\upsilon(\epsilon)$ in energy channel
$\epsilon$ with no scattering \citep{kru99}, we apply a linear fit
to $t_{sol}(\epsilon)=t_{AU}(\epsilon)-L/\upsilon(\epsilon)$, where
$L$ is the IMF path length from the acceleration site on the Sun to
the in-situ spacecraft, $t_{AU}(\epsilon)$ and $t_{sol}(\epsilon)$
are the rise time of electron flux at 1 AU and the release time at
the Sun in energy channel $\epsilon$, respectively. Then the
extrapolated electron solar release time $t_{sol}$ ($\epsilon
\rightarrow \infty$) is 05:19 UT $\pm$ 5 minutes, and the inferred
IMF path length $L$ is 1.3 $\pm$ 0.2 AU. To further confirm this
evaluation, adapting the solar wind speed of $\sim$ 380 km/s during
this event, the IMF path length is calculated by solution of the IMF
equation deduced from the solar wind model \citep{par58} to be
$\sim$ 1.19 AU. Taking $\upsilon(\epsilon)$ to be $\sim$ 0.5$c$ in
the ACE/EPAM energy channel $\epsilon$ of 60 keV, this channel leads
to a lower bound on the electron solar release time \citep{hag03}.
This is reasonable since electrons actually undergo interplanetary
scattering more or less, then the electron solar release time is
derived to be $\sim$ 05:20 UT. The results are consistent with the
previous evaluation.

To understand the anisotropy of the in-situ energetic electrons
requires an examination of particles transport direction relative to
the magnetic field. The Low Energy Foil Spectrometer oriented at
60$^{\circ}$ to the spin axis (LEFS60) of the ACE/EPAM provides
measurements of electron fluxes in 8 angular sectors, whose spatial
orientations projected onto the unit sphere are shown in the left
panel of Figure 2. The magnetic field vector at the peak time of
electron fluxes around 05:55 UT is obtained from the ACE/MAG in a
geocentric solar ecliptic (GSE) coordinate. Then the electron
pitch-angle distribution (PAD) is derived and shown in the right
panel of Figure 2. The letters correspond to the 8 angular sectors
in the 42 $-$ 65 keV channel. The distribution is normalized and
plotted against the pitch-cosine. Except for the uncertainty of
sector c, we characterize the strong anisotropic electrons as
beam-like when they are detected streaming mainly along the opposite
direction of the IMF. Therefore, the in-situ energetic electrons are
probably accelerated in a confined site on the Sun, e.g., flaring
AR, rather than a wide-spread CME-driven shock, and the open-field
lines streamed along by energetic electrons connect the spacecraft
to the acceleration region in a negative polarity.

\section{REMOTE-SENSING OBSERVATIONS}

To identify the solar source of the beam-like energetic electrons,
we first study the full-disk 195 ${\rm \AA}$ images from the
Extreme-Ultraviolet Imaging Telescope \citep{del95}, as well as the
WL observations from the Large Angle and Spectrometric Coronagraph
\citep{bru95} on board the Solar and Heliospheric Observatory
(SOHO). As shown in Figure 3, two western ARs were responsible for
the noticeable coronal activities in the time window of the electron
event: AR 10938, located at the far west limb, produced a faint CME
with an angular width of $67^{\circ}$ and a projection speed of 295
km/s. AR 10939, located at S06W63 and marked by the red box, 
produced an impulsive B7.3 flare and a jet which probably extended
to a few solar radius observed by the LASCO C2 coronagraph and
marked by the red arrow. By applying running difference to the C2
images, the WL jet signature propagates to southwest at an evaluated
speed of $\sim$ 190 km/s.

To further confirm the origin of the WL jet signature, EUV
observations with high temporal cadence and spatial resolution are
introduced. Figure 4 shows the development of a jet eruption from
the AR 10939. In panel (a), the 171 ${\rm \AA}$ image at the peak
time of the B7.3 flare is obtained from the Transition Region and
Coronal Explorer \citep{han99}. To enhance the spatial structure, we
apply a high-pass filter by subtracting a smoothed image from the
original one. The hard X-ray (HXR) source is reconstructed by the
Reuven Ramaty High Energy Solar Spectroscopic Imager \citep{lin02}
and co-aligned with the TRACE data. Red contour lines indicate the
25 $-$ 50 keV nonthermal bremsstrahlung source integrated from 05:16
to 05:17 UT. It is found the HXR source is located in a compact
flaring region where the jet erupted nearby to the eastern
large-scale loop structure. Panels (b) $-$ (d) consist of
running-difference images illustrating the jet evolution. The EUV
jet propagates to southwest at an evaluated speed of $\sim$ 205 km/s
during 05:16 to 05:25 UT. A simple estimation of the spatial and
temporal correlation suggests that the WL jet signature is the
counterpart of the EUV jet erupted from AR 10939. The velocity of
the jet slightly decreased at $\sim$ 3$R_{s}$.

The jet was also observed by the EUV Imaging Spectrometer
\citep{cul07} onboard the $Hinode$ satellite. A raster scan using
$1^{''}$ slit was performed with the EIS from 05:22 to 05:44 UT. The
jet signature is seen in the 255.95 $-$ 256.64 $\rm {\AA}$ intensity
map in the left panel of Figure 5. The waveband consists of the main
emission line He II peaked at 256.32 $\rm {\AA}$ and a blended one
Si X at 258.37 $\rm {\AA}$. We do a 2-component Gaussian fit and
decompose it into two intensity maps, respectively shown in the
right panel of Figure 5. The compact flaring region where the jet
erupted is shown in both of the two intensity maps. However, the jet
itself is only detected by the hot coronal line Si X rather than the
cool line He II, suggesting the jet is actually formed in a coronal
site.

The compact flaring region where the jet erupted is further analyzed
using the X-Ray Telescope \citep{gol07} onboard the $Hinode$
satellite. Figure 6 shows the XRT image in $\rm Al_{-}poly$ filter
whose temperature response peaks at 7 MK. It is, after co-alignment,
overlaid with the magnetic field polarities obtained from the
Michelson Doppler Imager \citep{sch95} onboard the SOHO. The compact
flaring region then is identified as a loop-like and cusp structure
straddled on the right inversion line. We also mark the footpoints
of open-field lines calculated using the potential-field
source-surface model \citep{sch03}. It is derived at 00:04 UT and
rotated to 05:22 UT. The open-field lines are rooted in an area of
negative polarity nearby to the jet site. Therefore the jet is
probably produced by the reconnection between closed loops straddled
on the middle inversion line and the western open-field lines.

\section{RELATING TO THE SOLAR SOURCE}

Based on the above analysis, the jet activity is expected to be
closely related to the in-situ energetic electrons. The magnetic
configuration, on the other hand, plays a crucial role in triggering
jet eruption and guiding energetic particles from the acceleration
site to interplanetary space. As shown in Figure 6, the magnetogram
in the vicinity of AR 10939 is quadrupolar and potentially favorable
for jet formation. It is, however, not possible to simulate a
precise coronal magnetic configuration at the solar limb since the
projection effect. The EIS velocity map, from another perspective,
gives us a clue to understand the magnetic topology. The redshifted
downflows observed in AR corresponding to closed loops are well
accepted. The blueshifted outflows, usually persist at the edges of
ARs, are variously interpreted as magnetic funnels playing a role in
coronal plasma circulation \citep{mar08}, expansion of large-scale
loops that lie over ARs \citep{har08}, and reconnection along QSLs
representing strong gradients of magnetic connectivity
\citep{bak09}.

In Figure 7, the EIS Fe XII emission line velocity map is contoured
with the co-aligned MDI magnetogram and sketched with a proposed
coronal magnetic topology. Doppler velocities are determined by
fitting a Gaussian function to the calibrated spectral profile.
Blueshifts (redshifts) correspond to negative (positive) Doppler
velocities in a range of -25 $-$ 25 km/s. Black contour lines
indicate positive longitudinal magnetic field, and green negative.
The strong blueshifted outflows might be explained by either of the
mechanisms aforementioned. The redshifted area corresponds to closed
loops of the AR 10939 marked by the red arcs. The jet is produced by
the reconnection between closed loops and negative open-field lines
predicted by the PFSS model and marked by the black solid lines,
along which accelerated particles escape into interplanetary space
be detected by near-Earth spacecraft. Note that the in-situ electron
PAD (shown in Figure 2) is consistent with the negative polarities
of open-field lines. A CME lifted off from the AR 10938 is marked by
the black dashed lines, which might compress the open-field lines to
facilitate reconnection with the closed loops and the jet eruption
(see discussion in Section 5).

Timing information gives us another clue to relate to the solar
source. Figure 8 shows the comparison of electron solar release, jet
duration, and the multi-wavelength flare emission. The electrons are
released at the Sun at 05:19 UT $\pm$ 5 minutes (see Section 2) and
marked by the orange bar. The EUV jet duration is lasting from 05:16
UT to 05:25 UT (see Section 3) and marked by the red bar. The
temporal profiles are shown by GOES $1-8 \ \rm {\AA}$ SXR, RHESSI 12
$-$ 50 keV HXR, and the RSTN/Learmonth 245 MHz radio emission,
respectively. The WIND/WAVES radio dynamic spectra in the frequency
range of 20 kHZ $-$ 14 MHz show clear type III bursts, which are
generated by electron beams as they propagate along magnetic field
lines from the solar corona to interplanetary medium. It is evident
that the electron solar release has a good temporal correlation with
the jet eruption, flare emission and the consequent type III radio
bursts, suggesting a physical link of magnetic reconnection, jet
production, energetic radiation, and the particle acceleration.

\section{DISCUSSION AND SUMMARY}

Table 1 summaries the impulsive electron events and west hemisphere 
solar activities observed on 2007 January 24. The well-connected AR 
10939 produced three jets identified in EUV images from the TRACE 
and STEREO/EUVI, respectively. The first jet erupted at $\sim$ 00:35 
UT was associated with a HXR flare detectable from the RHESSI and 
type III radio bursts from the WIND/WAVES. It is related to the 
electron event 1 whose solar release time is deduced at $\sim$ 00:37 
UT, and consistent with the jet eruption and flare emission. The 
second jet erupted at $\sim$ 01:15 UT without HXR or radio 
signature. No corresponding electron flux enhancement was recorded 
by the in-situ spacecraft. The third jet and the closely related 
electron event 2 are afore comprehensively analyzed.

The limb-side AR 10938 produced two CMEs (obtain from the LASCO CME
catalog: $http://cdaw.gsfc.nasa.gov/CME_{-}list/index.html$). The
first CME with an angular width of $67^{\circ}$ and a projection
speed of 295 km/s lifted off at $\sim$ 05:13 UT, temporally close to
the jet eruption at $\sim$ 05:16 UT and the electron solar release
at $\sim$ 05:19 UT. We interpret that this CME might either
facilitate the reconnection occurrence and jet eruption or further
accelerate particles in a higher coronal site via 3D reconnection
with the open fields. The second CME with a even larger angular
width of $72^{\circ}$ and a higher projection speed of 381 km/s
lifted off at $\sim$ 15:47 UT, however, was not temporally related
to a SEP event. Furthermore, no type II radio bursts were recorded 
in the time window of the impulsive electron events. Therefore, the 
jets in association with HXR flares and type III radio bursts, 
rather than the slow partial CMEs, are the crucial contributors to 
the interplanetary electron streams.

We combine a wide range of datasets, specifically in-situ electron
measurements and remote-sensing solar observations, to investigate
the acceleration source of the impulsive electron events. Data 
analysis show that: (1) The electron solar release has a good 
temporal correlation with the jet eruption and the multi-wavelength 
flare emission, namely SXR, HXR and type III radio burst. (2) The 
jet is originated from a well-connected AR whose magnetic polarity
structure favors the eruption, then formed in a coronal site, and
extended to a few solar radius. (3) The PFSS modeled open fields are 
rooted in negative polarities, along which energetic particles 
escape from the flaring AR to the near-Earth space, consistent with
the in-situ electron PAD. The results convincingly confirm the
viewpoint that the coronal jet is probably a key to understand the
production of interplanetary electron streams. However, many 
impulsive SEP events are not associated with jets or even lacking 
of X-ray signatures \citep{nit08,mas09}. Possibly the acceleration
takes place in high corona via 3D reconnection or other energize
processes. 

To be emphasized that the magnetic topology plays a crucial role in 
triggering jet eruption and guiding particles transport. Based on 
the velocity map, magnetogram and the potential-field extrapolation, 
we propose a coronal field configuration relating the impulsive SEP 
event to the solar source. Detailed reconstruction of magnetic 
fields, unavailable to the present study, should be further 
introduced to explore this topic.

\acknowledgments

We are very grateful to the WIND/3DP/WAVES, ACE/EPAM/MAG,
Hinode/EIS/XRT, SOHO/EIT/MDI/LASCO, STEREO/EUVI, RSTN/Learmonth,
RHESSI, and TRACE teams for providing the data used in this study.
This work was supported by the rolling grant from Science $\&$
Technology Facilities Council (STFC) of the UK.

\begin{figure}
\epsscale{.60} \plotone{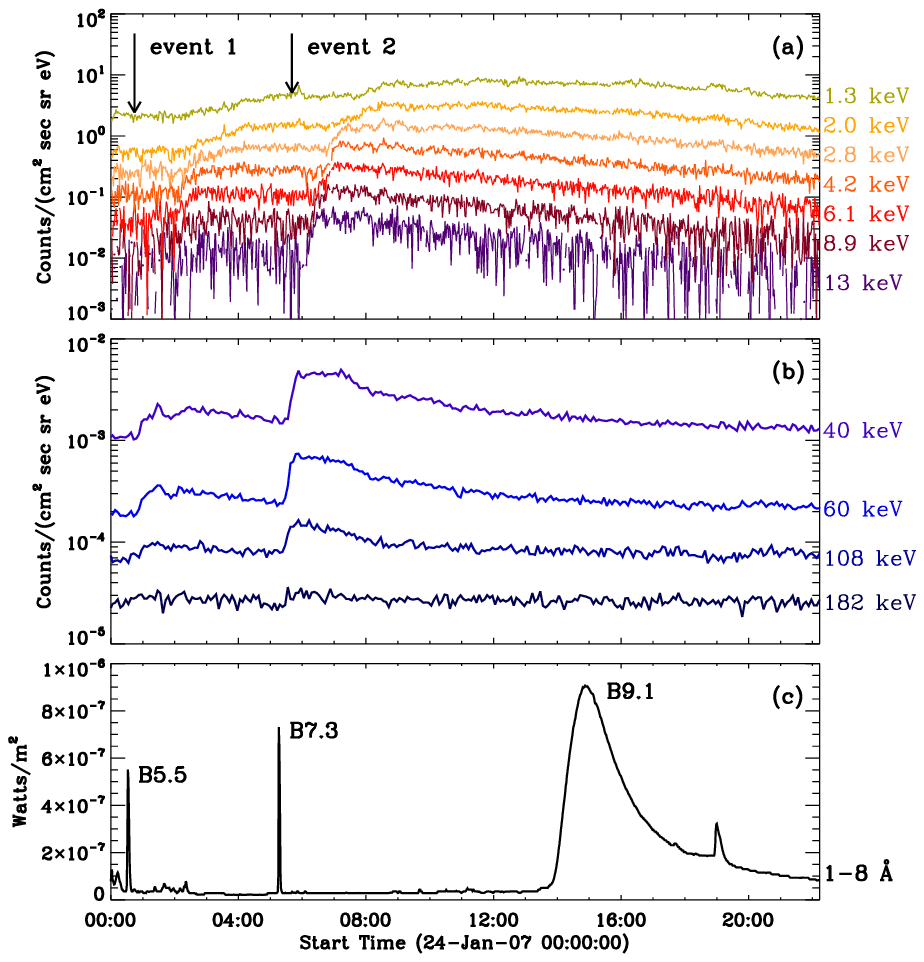} \caption{Impulsive electron events
observed on 2007 January 24. (a) Intensity profiles of in-situ
energetic electrons from $\sim$ 1 to 13 keV detected by the
WIND/3DP, and (b) from 38 to 315 keV by the ACE/EPAM, compared with
(c) the GOES SXR $1-8 \ \rm {\AA}$ light curve.
\label{fig1}}
\end{figure}

\begin{figure}
\epsscale{.80} \plotone{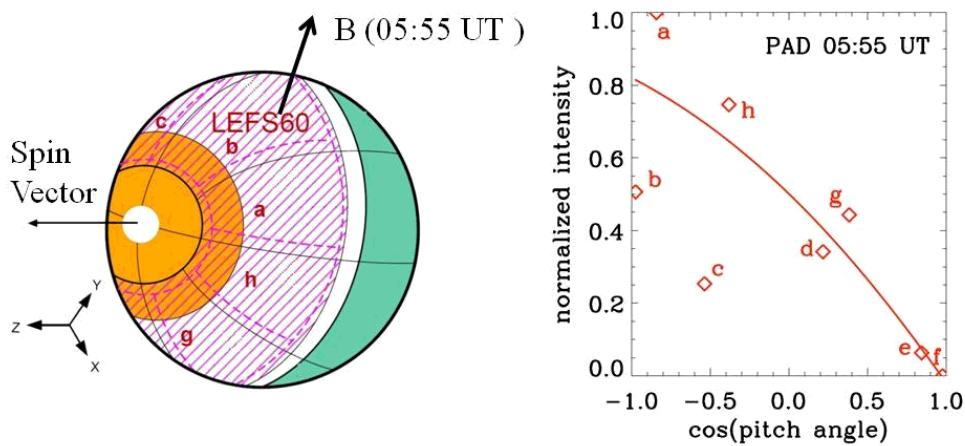} \caption{Left: Schematic view of
the ACE/EPAM with the LEFS60 projected onto a unit sphere (adapted
from Gold et al. 1998), and the magnetic field vector derived from
the ACE/MAG in a GSE coordinate. Right: The normalized PAD at the
peak time of the electron intensities in energy channel of 42 $-$
65 keV.
\label{fig2}}
\end{figure}

\begin{figure}
\epsscale{.50} \plotone{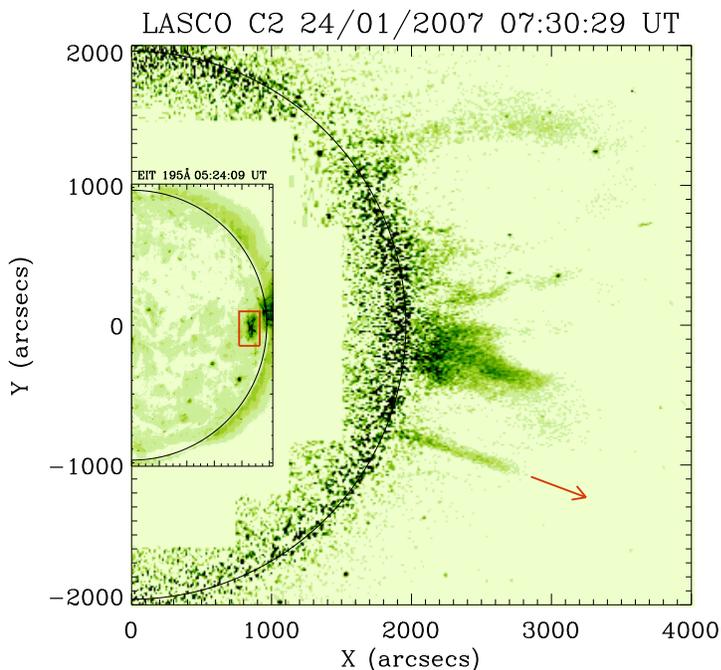} \caption{Composite data in WL and
EUV images obtained from the SOHO/LASCO C2 and the SOHO/EIT,
respectively. AR 10939 is marked by the red box, and a WL jet
signature by the red arrow. A faint CME is observed to be lifted
off from the nearby AR 10938.
\label{fig3}}
\end{figure}

\begin{figure}
\epsscale{.60} \plotone{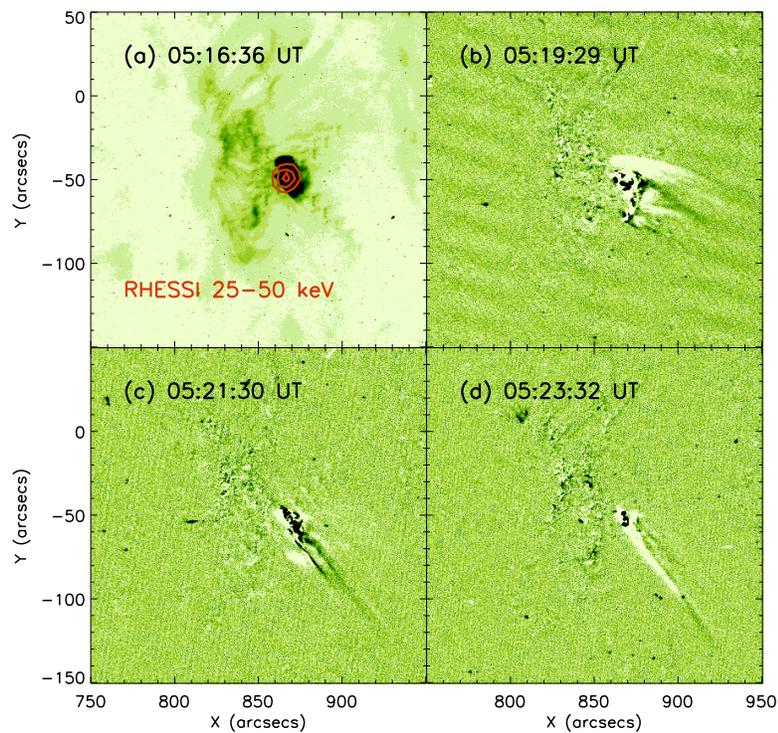} \caption{Jet eruption observed in
TRACE 171 ${\rm \AA}$ images. (a) High-pass filtered image at the
peak time of the B7.3 flare overlaid with the RHESSI HXR source in
the energy channel of 25 $-$ 50 keV. (b) $-$ (d) consist of
running-difference images illustrating the evolution of the jet.
\label{fig4}}
\end{figure}

\begin{figure}
\epsscale{.60} \plotone{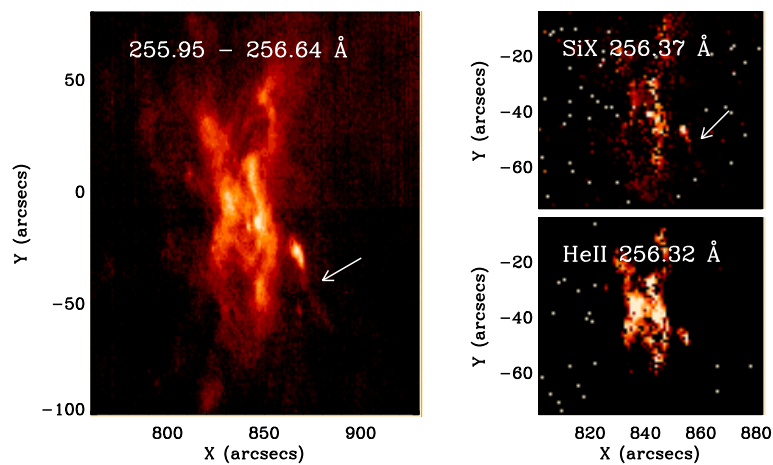} \caption{Jet signature observed in
Hinode/EIS 255.95 $-$ 256.64 $\rm {\AA}$ intensity map. After a
2-component Gaussian fit, the jet, marked by the white arrow, is
only detectable by a blended emission line Si X.
\label{fig5}}
\end{figure}

\begin{figure}
\epsscale{.60} \plotone{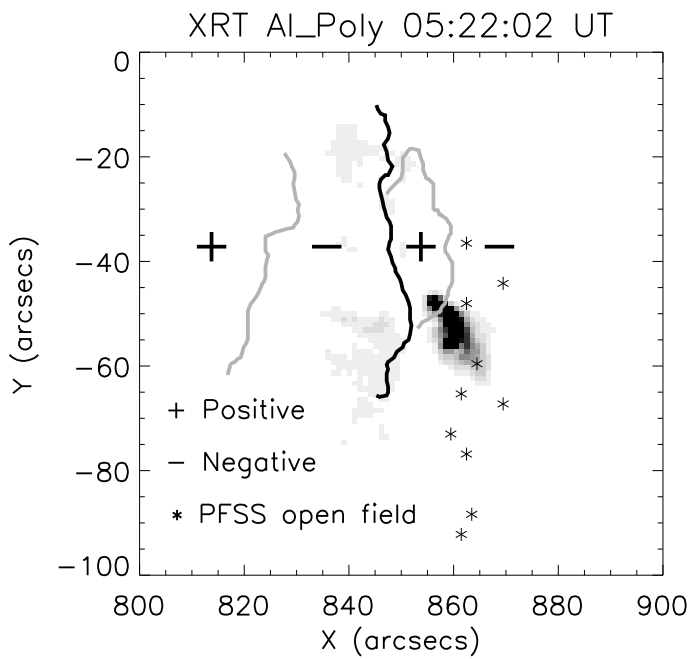} \caption{Hinode/XRT image by the
$\rm Al_{-}poly$ filter overlaid with the magnetic field polarities
obtained from the SOHO/MDI, and the footpoints of the PFSS modeled
open-field lines. Inversion lines are marked by the black and grey
lines. ``$+$" indicates positive longitudinal magnetic field, and
``$-$" negative. ``$*$" indicates footpoints of the open fields.
\label{fig6}}
\end{figure}

\begin{figure}
\epsscale{.80} \plotone{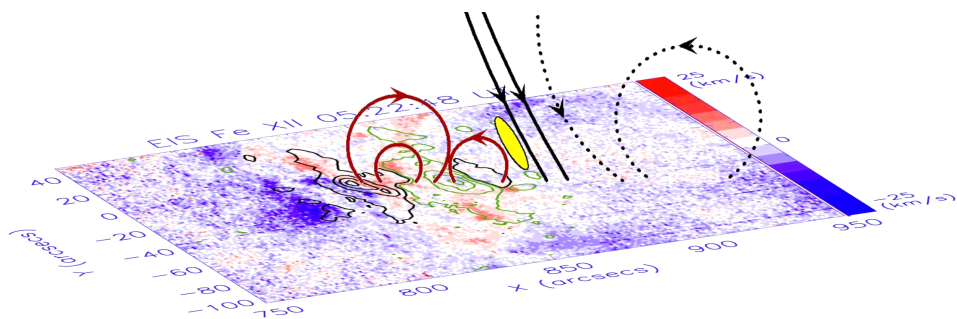} \caption{Hinode/EIS Fe XII
emission line velocity map contoured with the SOHO/MDI magnetogram,
and overlaid with a proposed coronal magnetic topology. Red arcs
indicate closed loops of the flaring AR 10939, black solid lines
indicate negative open-field lines. The reconnection site where the
jet erupted is marked by the yellow oval region. A CME lifted off
from the nearby AR 10938 is marked by the black dashed lines.
\label{fig7}}
\end{figure}

\begin{figure}
\epsscale{.50} \plotone{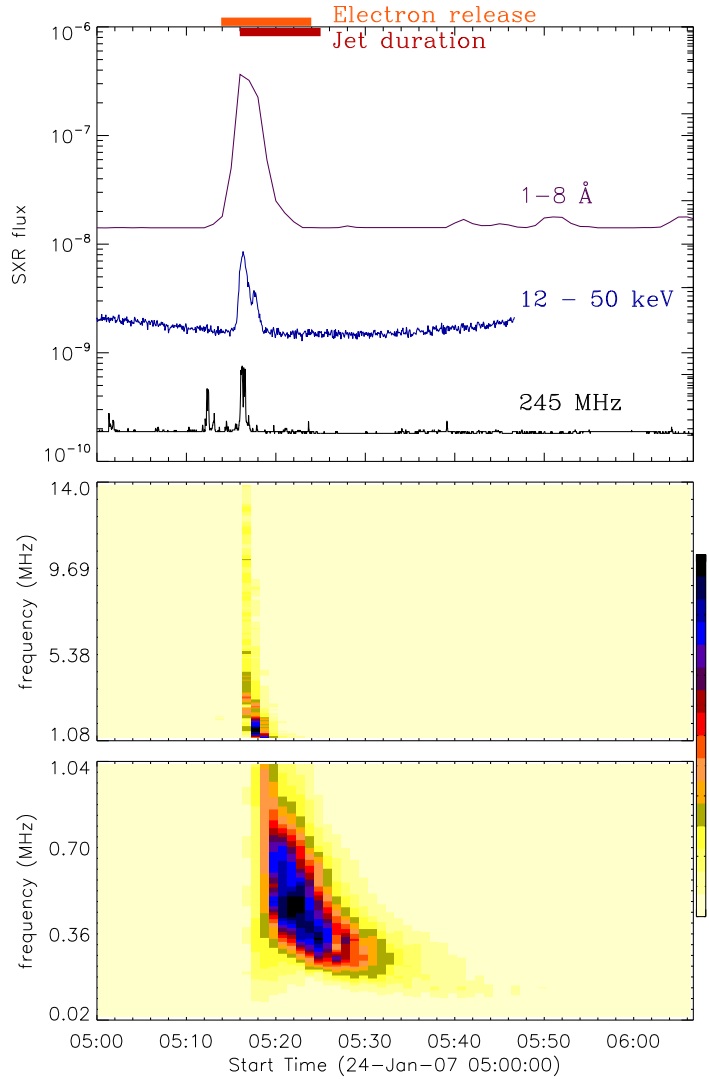} \caption{The electron solar release
compared with the jet eruption and multi-wavelength flare emission.
From top to bottom: Light curves of GOES SXR $1-8 \ \rm {\AA}$,
RHESSI HXR 12 $-$ 50 keV, RSTN/Learmonth 245 MHz radio emission, and
the WIND/WAVES radio spectrograms at frequency range of 20 kHz to 14
MHz.
\label{fig8}}
\end{figure}

\begin{deluxetable}{ccccccccc}
\tablecolumns{9}
\tablewidth{0pc}
\tablecaption{Electron events and west hemisphere solar activities
on 2007 January 24}

\tablehead{

\colhead{} & \colhead{} & \multicolumn{3}{c}{AR 10939} & \colhead{}
& \multicolumn{3}{c}{AR 10938} \\

\cline{3-5} \cline{7-9} \\

\colhead{SEP} & \colhead{\ \ \ \ \ } & \colhead{Jet} & \colhead{HXR
flare} & \colhead{Type III burst} & \colhead{\ \ \ \ \ } &
\multicolumn{3}{c}{CME} \\

\colhead{(UT)} & \colhead{} & \colhead{(UT)} & \colhead{} &
\colhead{} & \colhead{} & \colhead{liftoff} & \colhead{Width} &
\colhead{Speed} \\

\colhead{} & \colhead{} & \colhead{} & \colhead{} & \colhead{} &
\colhead{} & \colhead{(UT)} & \colhead{(deg)} & \colhead{(km/s)}

}

\startdata

00:37 & & $00:35^{a}$ & Y & Y & & \nodata & \nodata  &  \nodata \\

& & $01:15^{a}$ & N & N & & \nodata & \nodata & \nodata \\

& & -- & -- & -- & & 05:13 & 67 & 295 \\

05:19 & & $05:16^{b}$ & Y & Y & & \nodata & \nodata & \nodata \\

& & -- & -- & -- &  & 15:47 & 72 & 381 \\

\enddata

\tablenotetext{a}{Jets were identified by STEREO/EUVI in 10 minutes
temporal cadence. $^{b}$Jet was identified by TRACE in $\sim$ 1
minute cadence. --: No jet, HXR flare, or type III burst observed.
$\cdot\cdot\cdot$: No CME observed.}

\end{deluxetable}

\end{document}